\documentclass[aps,prd,amssymb,twocolumn,superscriptaddress,floatfix,nofootinbib,10pt]{revtex4-2}
\usepackage{graphicx}
\usepackage{dcolumn}
\usepackage{hyperref}
\usepackage{amsmath}
\usepackage{amsfonts}
\usepackage{booktabs}
\usepackage{amssymb}
\usepackage[usenames]{color}
\usepackage{lineno}
\usepackage{bm}
\usepackage{mathptmx}
\usepackage{multirow}
\usepackage{braket}
\usepackage{tikz}
\usepackage[compat=1.1.0]{tikz-feynman}

\tikzfeynmanset{arrow size=1}

\begin{document}

\preprint{APS/123-QED}

\title{Correlation function for the $p ~f_1(1285)$ interaction}
 
\author{P. Encarnación}
	\affiliation{Departamento de F\'{i}sica Teórica and IFIC, Centro Mixto Universidad de Valencia-CSIC, Institutos de Investigaci\'{o}n de Paterna, Aptdo. 22085, E-46071 Valencia, Spain}
 
\author{A. Feijoo}
	\affiliation{Instituto de F\'{i}sica Corpuscular, Centro Mixto Universidad de Valencia-CSIC, Institutos de Investigaci\'{o}n de Paterna, Aptdo. 22085, E-46071 Valencia, Spain}
 
\author{E. Oset}
	\affiliation{Departamento de F\'{i}sica Teórica and IFIC, Centro Mixto Universidad de Valencia-CSIC, Institutos de Investigaci\'{o}n de Paterna, Aptdo. 22085, E-46071 Valencia, Spain}
    \affiliation{Department of Physics, Guangxi Normal University, Guilin 541004, China}
 
\date{\today}

\begin{abstract}
 We have addressed here the  problem of calculating the correlation function of a stable particle with a resonance, in particular one resonance that qualifies as a molecular state of two components. The formalism used requires to evaluate the scattering matrix of the stable particle with the molecule, a nuclear problem which we address by means of the fixed center approximation. We have applied the method to the interaction of a proton with the $f_1(1285)$ resonance, presently under investigation by the ALICE collaboration, where the $f_1(1285)$ is taken as a $K^*\bar{K}-\bar{K}^*K$ molecule. We find that the $p ~ f_1(1285)$ interaction develops a resonance state below the $p ~ f_1(1285)$ threshold, which leads to a depletion in the $p ~ f_1(1285)$ correlation function for small values of the proton momentum. The discussion presented shows that these type of studies can provide much information on the nature of some resonances and the existence of three-body bound states involving mesons and baryons.

\end{abstract}

\maketitle


\section{Introduction}
\label{sec:introduction}

Femtoscopic correlation functions (CFs) are emerging as an important tool to learn about hadron dynamics \cite{Fabbietti:2020bfg}. By measuring pairs of particles in high energy p-p, p-A, or A-A collisions, and comparing production probabilities versus uncorrelated probabilities to find the same pair, one induces the low energy scattering parameters of pairs of particles. For a large amount of couples for which one has no access in scattering experiments. Experiments are regularly being conducted \cite{ALICE:2017jto,ALICE:2018ysd,ALICE:2019hdt,ALICE:2019eol,ALICE:2019buq,ALICE:2019gcn,ALICE:2020mfd,ALICE:2021szj,ALICE:2021cpv,ALICE:2022enj,STAR:2014dcy,STAR:2018uho,Fabbietti:2020bfg,Feijoo:2024bvn,ALICE:2022yyh} and theoretical papers follow the trend \cite{Morita:2014kza,Ohnishi:2016elb,Sarti:2023wlg,Morita:2016auo,Hatsuda:2017uxk,Mihaylov:2018rva,Haidenbauer:2018jvl,Morita:2019rph,Kamiya:2019uiw,Kamiya:2021hdb,Kamiya:2022thy,Liu:2023uly,Vidana:2023olz,Albaladejo:2023pzq,Liu:2023wfo,Liu:2022nec,Torres-Rincon:2023qll,Molina:2023jov,Molina:2023oeu,Li:2024tof,Liu:2024uxn,Encarnacion:2024jge}. Most theoretical works compare experimental CFs with theory, eventually using the experiment to determine free parameters of the theory. Lately, in what is called the inverse problem, the task of learning as much as possible about the interaction, in a way which is minimally model dependent, has caught up and has revealed the potential of CFs to learn more than just the scattering length and effective range, like information below threshold, with the possibility of there being bound states of the pairs. In this sense it is shown in \cite{Ikeno:2023ojl} that the measurement of the $DK$ and $D_s\eta$ correlation functions provides enough information to predict the existence of a $DK$ bound state, corresponding to the $D_{s0}^*(2317)$ resonance. Similarly, it is also shown in \cite{Albaladejo:2023wmv} that the knowledge of the $D^{*+}D^0$ and $D^{*0}D^+$ CFs allows us to determine the existence of a bound state below threshold corresponding to the $T_{cc}(3875)$ with quite a good precision. Moreover, as shown in \cite{Molina:2023jov}, the $N^*(1535)$ could be also be inferred from the knowledge of the  $K^0\Sigma^+$, $K^+\Sigma^0$, $K^+\Lambda$ and $\eta p$ CFs. A bound state of $BD$ nature is also induced from the $BD$ correlation function \cite{Li:2024tof}, and the $\Sigma(1430)$ state, predicted in \cite{Oller:2000fj,Jido:2003cb} and recently observed by the Belle Collaboration in \cite{Belle:2022ywa}, can also be predicted from the information contained in the $\bar{K}N$ and related coupled channels CFs. The test has an added value, because if these states were to be compact quark systems with no coupling to molecular components, they would have no influence in the CF of the pairs of particles, so the CFs have much to say about the nature of some resonances. More details and relationship to production experiments where invariant mass distributions are measured can be seen in \cite{Albaladejo:2024lam}. 

With the promising information for correlation functions of pairs of particles, work has begun by looking at three body CFs \cite{DelGrande:2021mju,ALICE:2022boj,ALICE:2023gxp,ALICE:2023bny,Garrido:2024pwi}. While certainly some information concerning three body dynamics can be obtained from these correlation functions, there are certainly obvious problems in the interpretation of the results because of the role played by the off-shell (unphysical) scattering amplitudes of the pairs involved \cite{Khemchandani:2008rk,MartinezTorres:2008gy}. Yet, we find very promising the CF of pairs formed by a stable particle and a resonance. Particularly, if these resonances correspond to some molecular state of two other particles, the correlation functions can definitely provide information on this molecular structure.  This is the purpose of the present work, for which we have selected one particular case, the $p~f_1(1285)$ CF. The choice is motivated by the fact that this measurement is under study by the ALICE Collaboration at present \cite{Otoninfo} (first steps in this direction can be found in \cite{ALICE:2024rjz}).

We take advantage here to elaborate further on the relationship of three body forces and off shell extrapolation of two body forces. To illustrate it we follow the work of Ref.~\cite{MartinezTorres:2007sr}. In the Faddeev equations to get three body amplitudes one finds a diagram like the one of Fig.~\ref{fig:intro1} left, where the line labeled $\vec{k}_{int}$ in the diagram is off shell. Working with chiral lagrangians for meson-baryon pairs and two meson baryon systems, the off shell $t_{13}$ amplitude of Fig.~\ref{fig:intro1} can be written as
\begin{equation}
t(\vec{k}_{int})=t_{on}+(\vec{k}_{int}^2-m^2)\alpha \, ,
\label{t_3body_diag}
\end{equation}
where, by definition, $t_{on}$ is the amplitude when $\vec{k}_{int}$ is on shell ($\vec{k}_{int}^2=m^2$). Note that the evaluation of the diagram requires the use of the propagator of the line $\vec{k}_{int}$, given by $(\vec{k}_{int}^2-m^2)^{-1}$. This factor cancels the $(\vec{k}_{int}^2-m^2)$ one present in Eq.~(\ref{t_3body_diag}) leading to a contact interacion, or equivalently a three body force, as shown in Fig.~\ref{fig:intro2}(a). This three body force is unphysical because the off shell extrapolation of the two body amplitudes is unphysical. If one were to use another model to generate the two body amplitudes, even providing the same on shell amplitude, it would have a different off shell extrapolation, thus generating a different three body force. It is usual to introduce an empirical three body force in three body calculations, and one of its effects is to cancel the unphysical three body forces generated by the off shell extrapolation of the two body amplitudes. It is thus clear that these empirical three body forces are model dependent. They depend on the model used to obtain the two body amplitudes, as it is well documented in \cite{Polyzou:1990hks}. The novel aspect of \cite{MartinezTorres:2007sr} is that using the same chiral lagrangians employed to generate the two body amplitudes of Eq.~(\ref{t_3body_diag}), and expanding the chiral field $U=\exp(i\sqrt{2}\phi/f)$ function, where $\phi$ are the meson fields and $f$ the pion decay constant, in terms of meson fields, a three body force stems from these chiral lagrangians, which we show in Fig.~\ref{fig:intro2}(b), and cancels exactly the three body force obtained before from the two body amplitudes. The three body amplitude derived from the chiral lagrangian has provided a magnitude that cancels the unphysical three body forces generated by the off shell two body amplitudes, rendering a result that is free of unphysical contributions. With this finding, and performing the calculation with chiral lagrangians, one has two methods to proceed:
\begin{enumerate}
   \item Calculate the Faddeev diagrams with the chiral lagrangians and all the three body terms.
   \item Evaluate the Faddeev diagrams using only the on shell part of the two body amplitudes and ignore the three body terms.
\end{enumerate} 
The two methods render identical results. Actually, the procedure of scenario (2) was used in \cite{MartinezTorres:2008kh}, employing experimental on shell amplitudes, to determine the interaction of the $\pi\pi N$ and coupled channel systems, providing a good description of the $N^*(1710)$ resonance. With the discussion given above, and evaluating the two body amplitudes with chiral lagrangians, as we shall do, we find the Fixed Center Approximations (FCA), which relies upon on shell amplitudes, as an appropiate tool to evaluate the three body amplitude needed in our approach.

\begin{figure}
    \centering
    \begin{tikzpicture}[scale=0.9, baseline={(current bounding box.center)}]
        \begin{feynman}
            \vertex (k1) at (-2,2) {$k_1$};
            \vertex (k2) at (-2,1) {$k_2$};
            \vertex (k3) at (-2,0) {$k_3$};
            \vertex (k1p) at (2,2) {$k_1'$};
            \vertex (k2p) at (2,1) {$k_2'$};
            \vertex (k3p) at (2,0) {$k_3'$};
            \vertex (v1) [dot, style={fill,scale=0.01}] at (-0.75,2) {};
            \vertex (v2) [dot, style={fill,scale=0.01}] at (-0.75,0) {};
            \vertex (v3) [dot, style={fill,scale=0.01}] at (0.75,1) {};
            \vertex (v4) [dot, style={fill,scale=0.01}] at (0.75,0) {};
            
            \diagram* {
                (k1) -- [draw, -] (v1) -- [draw, -] (k1p),
                (k2) -- [draw, -] (v3) -- [draw, -] (k2p),
                (k3) -- [draw, -] (v2) -- [draw, -, edge label=\(\vec{k}_{int}\)] (v4) -- [draw, -] (k3p),
                (v1) -- [draw, out=0, in=0, looseness=0.7] (v2),
                (v1) -- [draw, out=180, in=180, looseness=0.7] (v2),
                (v3) -- [draw, out=0, in=0, looseness=0.9] (v4),
                (v3) -- [draw, out=180, in=180, looseness=0.9] (v4)
            };
            
            \node at (-0.75,1.25) {\(t^2\)};
            \node at (0.75,0.5) {\(t^1\)};
            \node at (0,-0.75) {(a)};
            
        \end{feynman}
    \end{tikzpicture}
    \begin{tikzpicture}[scale=0.9, baseline={(current bounding box.center)}]
        \begin{feynman}
            \vertex (k1) at (-2,2) {$k_1$};
            \vertex (k2) at (-2,1) {$k_2$};
            \vertex (k3) at (-2,0) {$k_3$};
            \vertex (k1p) at (2,2) {$k_1'$};
            \vertex (k2p) at (2,1) {$k_2'$};
            \vertex (k3p) at (2,0) {$k_3'$};
            \vertex (v1) [dot, style={fill,scale=0.01}] at (-0.75,2) {};
            \vertex (v2) [dot, style={fill,scale=0.01}] at (-0.75,1) {};
            \vertex (v3) [dot, style={fill,scale=0.01}] at (0.75,1) {};
            \vertex (v4) [dot, style={fill,scale=0.01}] at (0.75,0) {};
            
            \diagram* {
                (k1) -- [draw, -] (v1) -- [draw, -] (k1p),
                (k3) -- [draw, -] (v4) -- [draw, -] (k3p),
                (k2) -- [draw, -] (v2) -- [draw, -, edge label=\(\vec{k}_{int}\)] (v3) -- [draw, -] (k2p),
                (v1) -- [draw, out=0, in=0, looseness=0.9] (v2),
                (v1) -- [draw, out=180, in=180, looseness=0.9] (v2),
                (v3) -- [draw, out=0, in=0, looseness=0.9] (v4),
                (v3) -- [draw, out=180, in=180, looseness=0.9] (v4)
            };
            
            \node at (-0.75,1.5) {\(t^3\)};
            \node at (0.75,0.5) {\(t^1\)};
            \node at (0,-0.75) {(b)};
            
        \end{feynman}
    \end{tikzpicture}
    \caption{Diagrammatic representation of the terms needed to compute three body amplitudes. Figure taken from Ref. \cite{MartinezTorres:2007sr}}
    \label{fig:intro1}
\end{figure}

\begin{figure}
    \centering
    \begin{tikzpicture}[scale=0.9, baseline={(current bounding box.center)}]
        \begin{feynman}
            \vertex (k1) at (-2,2) {$k_1$};
            \vertex (k2) at (-2,1) {$k_2$};
            \vertex (k3) at (-2,0) {$k_3$};
            \vertex (k1p) at (2,2) {$k_1'$};
            \vertex (k2p) at (2,1) {$k_2'$};
            \vertex (k3p) at (2,0) {$k_3'$};
            \vertex (v1) [dot, style={fill,scale=0.01}] at (-0.75,2) {};
            \vertex (v2) [dot, style={fill,scale=0.01}] at (0,0) {};
            \vertex (v3) [dot, style={fill,scale=0.01}] at (0.75,1) {};
            
            \diagram* {
                (k1) -- [draw, -] (v1) -- [draw, -] (k1p),
                (k2) -- [draw, -] (v3) -- [draw, -] (k2p),
                (k3) -- [draw, -] (v2) -- [draw, -] (k3p),
                (v1) -- [draw, out=0, in=0, looseness=0.3] (v2),
                (v1) -- [draw, out=180, in=180, looseness=0.3] (v2),
                (v3) -- [draw, out=0, in=0, looseness=0.4] (v2),
                (v3) -- [draw, out=180, in=180, looseness=0.4] (v2)
            };
            
            \node at (0,-0.75) {(a)};
            
        \end{feynman}
    \end{tikzpicture}
    \begin{tikzpicture}[scale=0.9, baseline={(current bounding box.center)}]
        \begin{feynman}
            \vertex (k1) at (-2,2) {$k_1$};
            \vertex (k2) at (-2,1) {$k_2$};
            \vertex (k3) at (-2,0) {$k_3$};
            \vertex (k1p) at (2,2) {$k_1'$};
            \vertex (k2p) at (2,1) {$k_2'$};
            \vertex (k3p) at (2,0) {$k_3'$};
            \vertex (v1) [dot, style={fill,scale=0.5}] at (0,0) {};
            
            \diagram* {
                (k1) -- [draw, -] (v1) -- [draw, -] (k1p),
                (k2) -- [draw, -] (v1) -- [draw, -] (k2p),
                (k3) -- [draw, -] (v1) -- [draw, -] (k3p)
            };
            
            \node at (0,-0.75) {(b)};
            
        \end{feynman}
    \end{tikzpicture}
    \caption{Diagrammatic illustration of the three body force origin (a) due to cancellation of the propagator in Fig. \ref{fig:intro1}(a) with the off-shell part of the chiral amplitude, (b) at the tree level. Figure taken from Ref. \cite{MartinezTorres:2007sr}.}
    \label{fig:intro2}
\end{figure}

The $f_1(1285)$ state appears in a natural way as a dynamically generated state from the interaction of $K^*\bar{K}-\bar{K}^*K$ in isospin $I=0$ \cite{Lutz:2003fm,Roca:2005nm,Garcia-Recio:2010enl,Zhou:2014ila,Geng:2015yta,Lu:2016nlp}.  Within this picture the decay of the $f_1(1285)$ to $a_0(980)\pi$ and $f_0(980)\pi$ has been investigated in \cite{Aceti:2015zva} and the decay into $K\bar{K}\pi$ in \cite{Aceti:2015pma}, both with good agreement with experiment. The molecular nature of the state has also been tested in different reactions, $K^-p \to f_1(1285) \Lambda$ \cite{Xie:2015wja}, $J/\psi \to \phi f_1(1285)$ in \cite{Xie:2015lta}, $B^0_s\to J/\psi f_1(1285)$ decay in \cite{Molina:2016pbg}, $\tau \to f_1(1285)\pi\nu_\tau$ decay in \cite{Oset:2018zgc} and $\bar{B}^0 \to J/\psi f_1(1285)$ decay in \cite{He:2021exv}.  With so much experimental information favoring the molecular picture of the $f_1(1285)$, we face the $p~f_1(1285)$ interaction in the context of CFs, anticipating us to the experimental results. In the framework of \cite{Vidana:2023olz} the CF is obtained from the scattering matrix of the two participating particles, in this case the scattering amplitude of $N$ interacting with the $f_1(1285)$.  We must then calculate the $N$ scattering amplitude with the molecular system of $K^*\bar{K}-\bar{K}^*K$. Since the system is bound by about $100$ MeV, we find it practical and sufficiently accurate to evaluate it using the FCA, which has been widely used in the literature \cite{Foldy:1945zz,Brueckner:1953zz,Brueckner:1953zza,Chand:1962ec,Barrett:1999cw,Deloff:1999gc,Kamalov:2000iy}, particularly in the study of pion and kaon scattering with deuterium. With so much binding of the $K^*\bar{K}$ components in the $f_1(1285)$, the main assumption of the FCA that the components of the cluster are not modified in the scattering of the external particle, seems well justified.

Details and practical use of the approach can be seen in the review paper \cite{MartinezTorres:2020hus} and in \cite{Roca:2010tf}, and a pedagogical description of the derivation and the implicit approximations made can be seen in Appendix~A of the paper \cite{Malabarba:2024hlv}.

\section{Formalism}

\subsection{The scattering matrix for $p~f_1(1285)$}

\begin{figure*}
    $
    \begin{array}{ccccccc}
        &
        \begin{tikzpicture}[scale=0.9, baseline={(current bounding box.center)}]
            \begin{feynman}
                \vertex (name1) at (0,-0.5) {$\frac{1}{\sqrt{2}}\left[ (K^*\bar{K})^{I=0} - (\bar{K}^*K)^{I=0} \right]$};
                \vertex (name2) at (0,3.25) {};
                \vertex (iL) at (-0.5,0) {};
                \vertex (iR) at (0.5,0) {};
                \vertex (iN) at (-1.5,0.1) {$N$};
                \vertex (fL) at (-0.5,3) {};
                \vertex (fR) at (0.5,3) {};
                \vertex (fN) at (-1.5,2.9) {$N$};
                \vertex (v1) [dot, style={fill,scale=0.01}] at (-0.5,1.5) {};
                \vertex (v2) [dot, style={fill,scale=0.01}] at (0.5,1.5) {};
                
                \diagram* {
                    (name1) -- [boson, opacity=0.0] (name2),
                    (iN) -- [dashed, fermion] (v1) -- [dashed, fermion] (fN),
                    (iL) -- [fermion] (v1) -- [fermion] (fL),
                    (iR) -- [fermion] (v2) -- [fermion] (fR)
                };
                
            \end{feynman}
        \end{tikzpicture}
        & + &
        \begin{tikzpicture}[scale=0.9, baseline={(current bounding box.center)}]
            \begin{feynman}
                \vertex (name1) at (0,-0.5) {$\frac{1}{\sqrt{2}}\left[ (K^*\bar{K})^{I=0} - (\bar{K}^*K)^{I=0} \right]$};
                \vertex (name2) at (0,3.25) {};
                \vertex (iL) at (-0.5,0) {};
                \vertex (iR) at (0.5,0) {};
                \vertex (iN) at (-1.5,0.1) {$N$};
                \vertex (fL) at (-0.5,3) {};
                \vertex (fR) at (0.5,3) {};
                \vertex (fN) at (1.5,2.9) {$N$};
                \vertex (v1) [dot, style={fill,scale=0.01}] at (-0.5,1.5) {};
                \vertex (v2) [dot, style={fill,scale=0.01}] at (0.5,1.5) {};
                
                \diagram* {
                    (name1) -- [boson, opacity=0.0] (name2),
                    (iN) -- [dashed, fermion] (v1) -- [scalar] (v2) -- [dashed, fermion] (fN),
                    (iL) -- [fermion] (v1) -- [fermion] (fL),
                    (iR) -- [fermion] (v2) -- [fermion] (fR)
                };
                
            \end{feynman}
        \end{tikzpicture}
        & + &
        \begin{tikzpicture}[scale=0.9, baseline={(current bounding box.center)}]
            \begin{feynman}
                \vertex (name1) at (0,-0.5) {$\frac{1}{\sqrt{2}}\left[ (K^*\bar{K})^{I=0} - (\bar{K}^*K)^{I=0} \right]$};
                \vertex (name2) at (0,3.25) {};
                \vertex (iL) at (-0.5,0) {};
                \vertex (iR) at (0.5,0) {};
                \vertex (iN) at (-1.5,0.1) {$N$};
                \vertex (fL) at (-0.5,3) {};
                \vertex (fR) at (0.5,3) {};
                \vertex (fN) at (-1.5,2.9) {$N$};
                \vertex (v1) [dot, style={fill,scale=0.01}] at (-0.5,1) {};
                \vertex (v2) [dot, style={fill,scale=0.01}] at (0.5,1.5) {};
                \vertex (v3) [dot, style={fill,scale=0.01}] at (-0.5,2) {};
                
                \diagram* {
                    (name1) -- [boson, opacity=0.0] (name2),
                    (iN) -- [dashed, fermion] (v1) -- [scalar] (v2) -- [scalar] (v3) -- [dashed, fermion] (fN),
                    (iL) -- [fermion] (v1) -- [fermion] (v3) -- [fermion] (fL),
                    (iR) -- [fermion] (v2) -- [fermion] (fR)
                };
                
            \end{feynman}
        \end{tikzpicture}
    & +\ \dots \\
        + &
        \begin{tikzpicture}[scale=0.9, baseline={(current bounding box.center)}]
            \begin{feynman}
                \vertex (name1) at (0,-0.5) {$\frac{1}{\sqrt{2}}\left[ (K^*\bar{K})^{I=0} - (\bar{K}^*K)^{I=0} \right]$};
                \vertex (name2) at (0,3.25) {};
                \vertex (iL) at (-0.5,0) {};
                \vertex (iR) at (0.5,0) {};
                \vertex (iN) at (1.5,0.1) {$N$};
                \vertex (fL) at (-0.5,3) {};
                \vertex (fR) at (0.5,3) {};
                \vertex (fN) at (1.5,2.9) {$N$};
                \vertex (v1) [dot, style={fill,scale=0.01}] at (-0.5,1.5) {};
                \vertex (v2) [dot, style={fill,scale=0.01}] at (0.5,1.5) {};

                \diagram* {
                    (name1) -- [boson, opacity=0.0] (name2),
                    (iN) -- [dashed, fermion] (v2) -- [dashed, fermion] (fN),
                    (iL) -- [fermion] (v1) -- [fermion] (fL),
                    (iR) -- [fermion] (v2) -- [fermion] (fR)
                };
            \end{feynman}
        \end{tikzpicture}
        & + &
        \begin{tikzpicture}[scale=0.9, baseline={(current bounding box.center)}]
            \begin{feynman}
                \vertex (name1) at (0,-0.5) {$\frac{1}{\sqrt{2}}\left[ (K^*\bar{K})^{I=0} - (\bar{K}^*K)^{I=0} \right]$};
                \vertex (name2) at (0,3.25) {};
                \vertex (iL) at (-0.5,0) {};
                \vertex (iR) at (0.5,0) {};
                \vertex (iN) at (1.5,0.1) {$N$};
                \vertex (fL) at (-0.5,3) {};
                \vertex (fR) at (0.5,3) {};
                \vertex (fN) at (-1.5,2.9) {$N$};
                \vertex (v1) [dot, style={fill,scale=0.01}] at (-0.5,1.5) {};
                \vertex (v2) [dot, style={fill,scale=0.01}] at (0.5,1.5) {};

                \diagram* {
                    (name1) -- [boson, opacity=0.0] (name2),
                    (iN) -- [dashed, fermion] (v2) -- [scalar] (v1) -- [dashed, fermion] (fN),
                    (iL) -- [fermion] (v1) -- [fermion] (fL),
                    (iR) -- [fermion] (v2) -- [fermion] (fR)
                };
            \end{feynman}
        \end{tikzpicture}
        & + &
        \begin{tikzpicture}[scale=0.9, baseline={(current bounding box.center)}]
            \begin{feynman}
                \vertex (name1) at (0,-0.5) {$\frac{1}{\sqrt{2}}\left[ (K^*\bar{K})^{I=0} - (\bar{K}^*K)^{I=0} \right]$};
                \vertex (name2) at (0,3.25) {};
                \vertex (iL) at (-0.5,0) {};
                \vertex (iR) at (0.5,0) {};
                \vertex (iN) at (1.5,0.1) {$N$};
                \vertex (fL) at (-0.5,3) {};
                \vertex (fR) at (0.5,3) {};
                \vertex (fN) at (1.5,2.9) {$N$};
                \vertex (v1) [dot, style={fill,scale=0.01}] at (-0.5,1.5) {};
                \vertex (v2) [dot, style={fill,scale=0.01}] at (0.5,1) {};
                \vertex (v3) [dot, style={fill,scale=0.01}] at (0.5,2) {};

                \diagram* {
                    (name1) -- [boson, opacity=0.0] (name2),
                    (iN) -- [dashed, fermion] (v2) -- [scalar] (v1) -- [scalar] (v3) -- [dashed, fermion] (fN),
                    (iL) -- [fermion] (v1) -- [fermion] (fL),
                    (iR) -- [fermion] (v2) -- [fermion] (v3) -- [fermion] (fR)
                };
            \end{feynman}
        \end{tikzpicture}
    & +\ \dots
    \end{array}
    $
    \caption{Diagrams in the FCA for $Nf_1(1285)$ scattering}
    \label{fig:diagrams}
\end{figure*}

Within the chiral unitary approach the $f_1(1285)$ is a molecular state of $K^*\bar{K}-\bar{K}^*K$ \cite{Roca:2005nm}. In terms of these components one has
\begin{eqnarray}
    f_1(1285) &=& \frac{1}{\sqrt{2}}\left[ (K^*\bar{K})^{I=0} - (\bar{K}^*K)^{I=0} \right] \nonumber \\
    &=& \frac{1}{2}(K^{*+}K^- + K^{*0}\bar{K}^0 - K^{*-}K^+ - \bar{K}^{*0}K^0)
\end{eqnarray}
with the phases of the isospin multiplets $(K^+,K^0)$, $(\bar{K}^0,-K^-)$, $(K^{*+},K^{*0})$, $(\bar{K}^{*0},-K^{*-})$. The interaction of $K^*\bar{K}$ is in s-wave and the state emerges as $I^G(J^{PC})=0^+(1^{++})$.

In the FCA approach the scattering of a nucleon with the $K^*\bar{K}-\bar{K}^*K$ cluster is given diagramatically by the diagrams in Fig.~\ref{fig:diagrams}.

The FCA assumes that the cluster of $K^*\bar{K}-\bar{K}^*K$ remains unchanged through the multiple rescattering of the nucleon with its components. In Fig.~\ref{fig:diagrams} one separates the diagrams where the $N$ interacts first with the first particle of the cluster ($K^*$ or $\bar{K}^*$), and we call this sum the partition $T_1$. Similarly, the partition $T_2$ sums all the diagrams where the $N$ interacts first with the second particle ($K$ or $\bar{K}$). Then we have two coupled equations
\begin{eqnarray}
    T &=& T_1+T_2 \nonumber \\
    T_1 &=& t_1+t_1\bar{G}T_2 \nonumber \\
    T_2 &=& t_2 + t_2\bar{G}T_1
    \label{eq:FC_system}
\end{eqnarray}
where $\bar{G}$ is the nucleon propagator folded with the wave function of the cluster, and $t_1,t_2$ are the scattering matrices of $N$ with $K^*(\bar{K}^*)$ or $K(\bar{K})$ respectively (see Appendix \ref{ap:apA}). We have to make two considerations. The first one is the isospin that comes from the nucleon $N$ with the $K^*\bar{K}-\bar{K}^*K$ cluster. The second one is that we want to evaluate the scattering matrix of $Nf_1(1285)$ with the field normalization of the particles, which are different for $NK^*$, $N\bar{K}$ and for $Nf_1(1285)$ and meson-meson in the Mandl and Shaw normalization that we follow \cite{Mandl2007-xg}. We refrain for detailing that here, which is done in \cite{Roca:2010tf,Xiao:2011rc,Zhai:2024luy}. The consideration of the isospin leads to 
\begin{eqnarray}
    t_1 &=& \frac{3}{4} t_{pK^*}^{(1)} + \frac{1}{4} t_{pK^*}^{(0)} \nonumber \\
    t_1' &=& \frac{3}{4} t_{p\bar{K}^*}^{(1)} + \frac{1}{4} t_{p\bar{K}^*}^{(0)} \nonumber \\
    t_2 &=& \frac{3}{4} t_{p\bar{K}}^{(1)} + \frac{1}{4} t_{p\bar{K}}^{(0)} \nonumber \\
    t_2' &=& \frac{3}{4} t_{pK}^{(1)} + \frac{1}{4} t_{pK}^{(0)} 
\end{eqnarray}
where the superindex in $t^{(I)}$ stands for the total isospin of the nucleon with each component. The $t_i$ amplitudes without prime stand for the $K^*\bar{K}$ component, and those with primes for the $\bar{K}^*K$ one. Since the $pK^*$ does not mix with $p\bar{K}^*$ one has then
\begin{eqnarray}
    t_{1,av} = \frac{1}{2}(t_1+t_1') \nonumber \\
    t_{2,av} = \frac{1}{2}(t_2+t_2')
\end{eqnarray}

As for the different field normalizations, the changes to be made are simple,
\begin{equation}
    t_1\to\tilde{t}_1=\frac{M_c}{M_{K^*}}t_1 \hspace{0.2cm} ; \hspace{0.2cm} t_2\to\tilde{t}_2=\frac{M_c}{M_{K}}t_2
\end{equation}
and the same for $t_1'$ and $t_2'$. The $\bar{G}$ function then becomes $\tilde{G}$, given by
\begin{equation}
    \tilde{G} = \frac{1}{2M_c} \int \frac{d^3q}{(2\pi)^3} F_c(q)\frac{M_N}{E_N(q)} \frac{1}{q^0-E_N(q)+i\epsilon}
    \label{eq:Gtilde}
\end{equation}
where $M_c$ is the mass of the cluster ($f_1(1285)$), $M_N$ and $E_N(q)$ the nucleon mass and energy, and $q^0$ the energy carried by the $N$ in the rest frame of the cluster, given by
\begin{equation}
    q^0=\frac{s-m_p^2-M_c^2}{2M_c},
    \label{eq:q0}
\end{equation}
with $s$ the square of the total $Nf_1(1285)$ energy in the $Nf_1(1285)$ rest frame, and $m_p$ the proton (nucleon in general) mass.

The function $F_c(q)$ in Eq.~(\ref{eq:Gtilde}) is the form factor of the cluster normalized to 1 at $q=0$. With the normalization of fields suited to the $Nf_1(1285)$, Eqs.~(\ref{eq:FC_system}) become
\begin{eqnarray}
    \tilde{T} &=& \tilde{T}_1+\tilde{T}_2 \nonumber \\
    \tilde{T}_1 &=& \tilde{t}_{1,av}+\tilde{t}_{1,av}\tilde{G}\tilde{T}_2 \nonumber \\
    \tilde{T}_2 &=& \tilde{t}_{2,av} + \tilde{t}_{2,av}\tilde{G}T_1
\end{eqnarray}
and solving these equations one finds our final formula
\begin{equation}
    \tilde{T} = \frac{\tilde{t}_{1,av}+\tilde{t}_{2,av}+2\tilde{t}_{1,av}\tilde{t}_{2,av}\tilde{G}}{1-\tilde{t}_{1,av}\tilde{t}_{2,av}\tilde{G}^2}
    \label{eq:Ttotal}
\end{equation}

The form factor is readily obtained with the formalism inherent to the chiral unitary approach with a cutoff regularization of the loop. It is given by \cite{Roca:2010tf}
\begin{equation}
    F_c(q)=\tilde{F}_c(q)/\tilde{F}_c(q=0)
\end{equation}
with
\begin{eqnarray}
    \tilde{F}_c(q) = &\displaystyle\int&\frac{d^3p}{(2\pi)^3}\frac{1}{M_c-\omega_{K^*}(p)-\omega_{\bar{K}}(p)} \nonumber \\
    &\times& \frac{1}{M_c-\omega_{K^*}(\vec{p}-\vec{q})-\omega_{\bar{K}}(\vec{p}-\vec{q})} \nonumber \\
    &|\vec{p}|&<q_{max} \nonumber \\
    &|\vec{p}-\vec{q}|&<q_{max}
    \label{form_factor}
\end{eqnarray}
where $q_{max}$ is the cutoff momentum needed to regularize the loops in the chiral unitary approach, chosen to be $q_{max}=1000$ MeV, ensuring that the $f_1(1285)$ state is generated at its experimental mass.

We should note that the vector-baryon amplitudes carry a factor $\vec{\epsilon}\vec{\epsilon}'$ for $V(\epsilon)N\to V(\epsilon')N$. This factor is repeated in all the diagrams and factorizes at the end, since we have $\epsilon_i\epsilon_i''\epsilon_j''\epsilon_j'$, which using $\sum_{pol}\epsilon_i''\epsilon_j''=\delta_{ij}$ leads to $\epsilon_i\epsilon_i'$, and so on.

There is still one more element needed: the argument of the $t_i^{(I)}$ scattering amplitudes. Following \cite{Zhai:2024luy,Montesinos:2024eoy}, splitting the binding of the $f_1(1285)$ proportional to the $K^*,K$ masses one finds
\begin{eqnarray}
    s_1(NK^*)&=&m_N^2+(\xi m_{K^*})^2 + 2\xi m_{K^*} q^0 \nonumber \\
    s_2(N\bar{K})&=&m_N^2+(\xi m_{\bar{K}})^2 + 2\xi m_{\bar{K}} q^0
\end{eqnarray}
with $\xi=M_c/(m_{K^*}+m_{\bar{K}})$, and $q^0$ given by Eq.~(\ref{eq:q0}).

\subsection{Elastic unitarity of the $\tilde{T}$ matrix}

We face here the issue of elastic unitarity of the $\tilde{T}$ matrix that we have obtained. We start from Eq.~(\ref{eq:Ttotal}), with $\tilde{G}$ given by Eq.~(\ref{eq:Gtilde}). We find easily that
\begin{equation}
    \text{Im}\tilde{G} = -\frac{1}{2M_c}\frac{1}{2\pi}q F_c(q) M_N,
\end{equation}
with $q$ the momentum of the nucleon in the rest frame of the cluster. We recall that, relative to $q_{cm}$ in the nucleon-cluster rest frame, we have
\begin{equation}
    \frac{q}{q_{cm}} = \frac{\sqrt{s}}{M_c}
    \label{eq:qqcm}
\end{equation}
The rest of magnitudes in $\tilde{T}^{-1}$, $\tilde{t}_1,\tilde{t}_2$ and $\text{Re}\tilde{G}$, are functions of $\sqrt{s}$ and hence of $q^2$. We remove here the subindex $av$ for simplicity in the notation. Thus, $\text{Im}\tilde{G}$ is the only term linear in $q$ in $\tilde{T}$. We can write for convenience the identity
\begin{equation}
    \tilde{T}^{-1} = \left(\frac{1}{4}\frac{\tilde{t}_1+\tilde{t}_2}{\tilde{t}_1\tilde{t}_2} - \frac{1}{2}\tilde{G} \right) - \frac{1}{4}\frac{(\tilde{t}_1-\tilde{t}_2)^2}{\tilde{t}_1\tilde{t}_2}\frac{1}{\tilde{t}_1+\tilde{t}_2+2\tilde{t}_1\tilde{t}_2\tilde{G}}
\end{equation}
which allows us to make an expansion around the threshold in terms of $q$, keeping constant and linear terms in $q$. We find that the term linear in $q$ of $\tilde{T}^{-1}$, which means linear in $\text{Im}\tilde{G}$, is given by $-iA\text{Im}\tilde{G}$, with $A$ given by
\begin{equation}
    A = \left( \frac{1}{2} - \frac{1}{2}\frac{(\tilde{t}_1-\tilde{t}_2)^2}{(\tilde{t}_1+\tilde{t}_2+2\tilde{t}_1\tilde{t}_2\text{Re}\tilde{G})^2} \right)
\end{equation}

Now we recall our normalization of the $\tilde{T}$ matrix related to the ordinary amplitue of Quantum Mechanics, $f^{QM}$,
\begin{eqnarray}
    && \tilde{T} = -\frac{8\pi\sqrt{s}}{2M_N}f^{QM}\hspace{0.2cm} ; \nonumber \\
    && -\frac{8\pi\sqrt{s}}{2M_N}\tilde{T}^{-1} = (f^{QM})^{-1} \simeq -\frac{1}{a_0}+\frac{1}{2}r_0q_{cm}^2 - i q_{cm} 
    \label{eq:effectiverangeexpansion}
\end{eqnarray}
and equating the linear terms in $q$ in the middle equation we find
\begin{equation}
    -\frac{8\pi\sqrt{s}}{2M_N}(-A)(-\frac{1}{2M_c}\frac{1}{2\pi} q F_c(0) M_N) \equiv -q_{cm}
\end{equation}
which means, using Eq.~(\ref{eq:qqcm}),
\begin{equation}
    \frac{s}{M_c^2}A\bigg|_{th}q_{cm} \equiv q_{cm},
\end{equation}
where $\frac{s}{M_c^2}A|_{th}$ has to be evaluated at the threshold of the $p~f_1(1285)$ system.

We would wish that $\frac{s}{M_c^2}A|_{th}$ should be exactly 1, but this is not the case due to the different static approximations made in the derivation of $\tilde{T}$ (see Appendix A of the work \cite{Malabarba:2024hlv}). However, we expect it to be of the order of 1. In the present case we find it to be $1.52$\footnote{The nucleon propagator in Eq.~(\ref{eq:Gtilde}) is taken as the positive energy part of a full propagator, and $q_0$, $E_N(q)$ are taken in the cluster at rest frame, where the cluster form factor of  Eq.~(\ref{form_factor}) is evaluated. The actual value of the correction factor is complex, but the imaginary part is only 1\% of the real part.}. If we divide $\tilde{T}^{-1}$ by $\frac{s}{M_c^2}A|_{th}$, or what is the same, multiply $\tilde{T}$ by $\frac{s}{M_c^2}A|_{th}$, then the elastic unitarity is restored exactly, and this is what we do. Thus, we replace
\begin{equation}
    \tilde{T} \to \tilde{T}_{uni} = \frac{s}{M_c^2}A\bigg|_{th} \tilde{T}
\end{equation}
Note that this does not change the structure of $\tilde{T}$ below threshold, that we shall see below, and thus the position and width of the obtained peak do not change. Then we can use $\tilde{T}_{uni}$ to write the effective range expansion and obtain the scattering length and effective range from Eq.~(\ref{eq:effectiverangeexpansion}), with $\tilde{T}_{uni}$, and, thus,
\begin{eqnarray}
    \frac{1}{a_0} &=& \frac{8\pi\sqrt{s}}{2M_N} (\tilde{T}_{uni}^{-1})\bigg|_{th} \nonumber \\
    r_0 &=& \frac{1}{\mu}\left[ \frac{\partial}{\partial\sqrt{s}} \left( -\frac{8\pi\sqrt{s}}{2M_N}(\tilde{T}_{uni}^{-1})+iq_{cm} \right) \right]_{th}
    \label{eq:a0r0}
\end{eqnarray}
with $\mu$ the reduced mass of $p$ and $f_1(1285)$.

\subsection{Correlation function}

We follow closely the formalism of \cite{Vidana:2023olz}, with the normalization of fields of meson-baryon. The correlation function is given by
\begin{eqnarray}
    C_{p~f_1}(p) = 1 &+& 4\pi \int_0^\infty dr r^2 S_{12}(r) \theta(q'_{max}-|\vec{p}|) \nonumber \\
    &&\{|j_0(pr)+\tilde{T}(\sqrt{s})\tilde{G}(s,r)|^2-j_0^2(pr)\}
    \label{eq:CF}
\end{eqnarray}
with $j_0(x)$ the spherical Bessel function, and $S_{12}(r)$ the source function for the pair production, which is usually parametrised as a Gaussian, 
\begin{equation}
    S_{12}(r)=\frac{1}{(4\pi R^2)^{3/2}}\exp(-r^2/4R^2),
    \label{eq:source}
\end{equation}  
with $R$ a parameter that is of the order of $1$ fm in pp collisions and $5$ fm in heavy-ion collisions. The value of $R$ is not known a priori, but it is very rewarding to see that from the works addresing the inverse problem of getting information from the data in a model independent way \cite{Ikeno:2023ojl,Albaladejo:2023wmv,Molina:2023jov,Li:2024tof}, the value of $R$ can be induced from fits to the data with very good precision. We shall evaluate the correlation function for several values of $R$. 

The function $\tilde{G}(s,r)$ is given by
\begin{eqnarray}
    \tilde{G}(s,r) && =\int \frac{d^3q}{(2\pi)^3} \frac{1}{2\omega_{f_1}(q)} \frac{M_N}{E_N(q)} \frac{j_0(qr)}{\sqrt{s}-\omega_{f1}(q)-E_N(q)+i\epsilon} \nonumber \\
    |\vec{q}| && < q'_{max}
    \label{eq:loopCF}
\end{eqnarray}   
The cutoff $q'_{max}$ in Eq.~(\ref{eq:loopCF}) is not necessarily the same $q_{max}$ used for the $f_1(1285)$ wave function, but should be of the same order of magnitude. Yet, the presence of the $j_0(qr)$ factor in $\tilde{G}(s,r)$ provides a convergence factor that makes the dependence on $q'_{max}$ very smooth. In this work we choose $q'_{max}=630$ MeV, and changes to $q'_{max}=800$ MeV result in changes in the CF of the level of 3\%.

\section{Results}

\begin{figure}
    \centering
    \includegraphics[width=\linewidth]{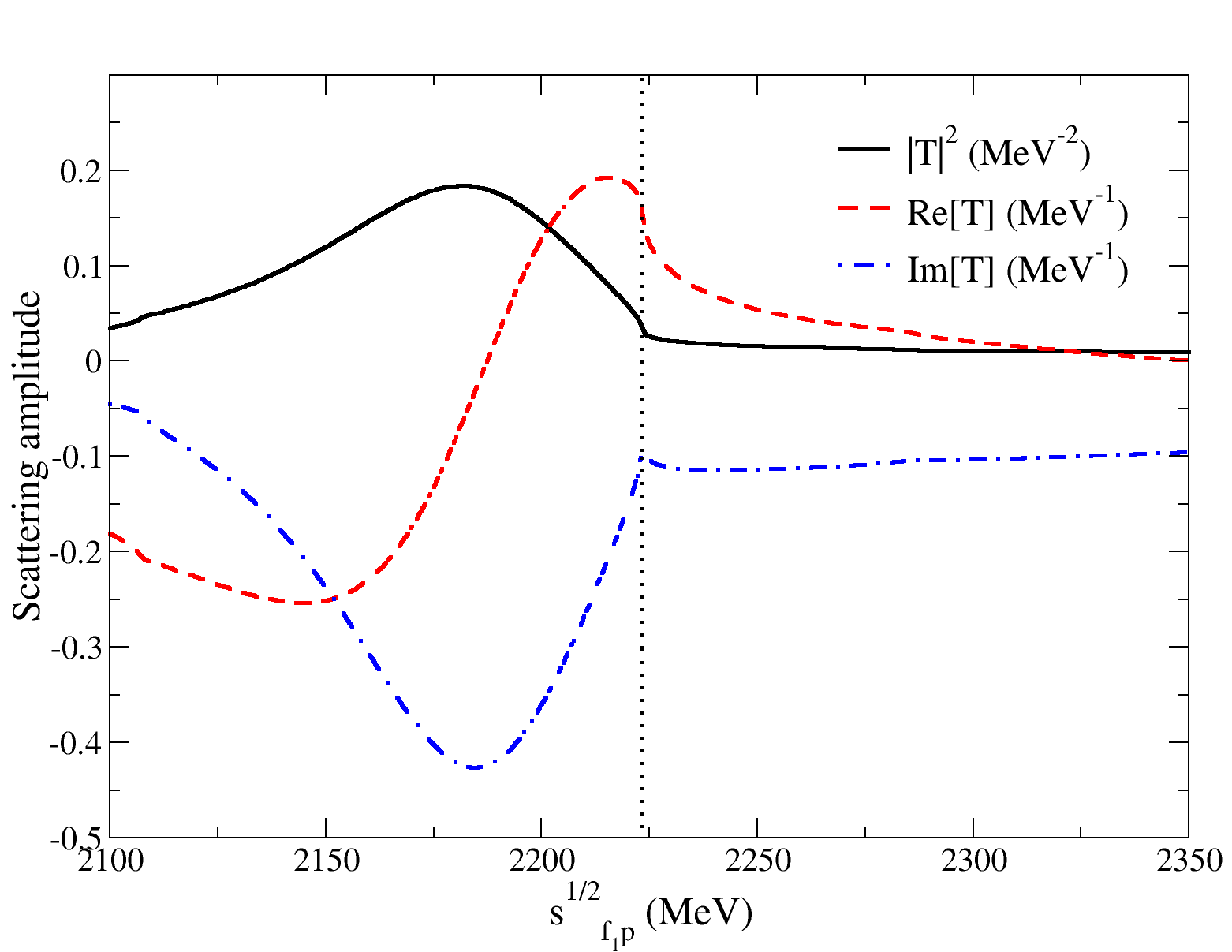}
    \caption{$\text{Re}\tilde{T}$, $\text{Im}\tilde{T}$ and $|\tilde{T}|^2$ for the $p~f_1(1285)$ interaction as a function of $\sqrt{s}$.}
    \label{fig:amplitude}
\end{figure}

In the first place we show the results for the $\tilde{T}$ matrix for the scattering of $p~f_1(1285)$ in Fig.~\ref{fig:amplitude}, where we show $\text{Re}\tilde{T}$, $\text{Im}\tilde{T}$ and $|\tilde{T}|^2$. We see there that we find a peak below the $p~f_1(1285)$ threshold. The squared absolute value of the amplitude peaks around $2180$ MeV, which is about $40$ MeV below the $p~f_1(1285)$ threshold. We associate this to the bound state produced by the interaction of the $p$ with the $\bar{K}^*,K^*,K,\bar{K}$ components of the cluster. It is interesting to see that $\text{Im}\tilde{T}$ has a peak and $\text{Re}\tilde{T}$ changes from negative to positive around that same energy, as it occurs with an ordinary resonance. From $|\tilde{T}|^2$ one can extract the position and width of the peak, obtaining
\begin{equation}
    M_{Nf_1} = 2182\text{ MeV} \hspace{1cm} \Gamma_{Nf_1} = 78\text{ MeV}
\end{equation}

\begin{figure}
    \centering
    \includegraphics[width=\linewidth]{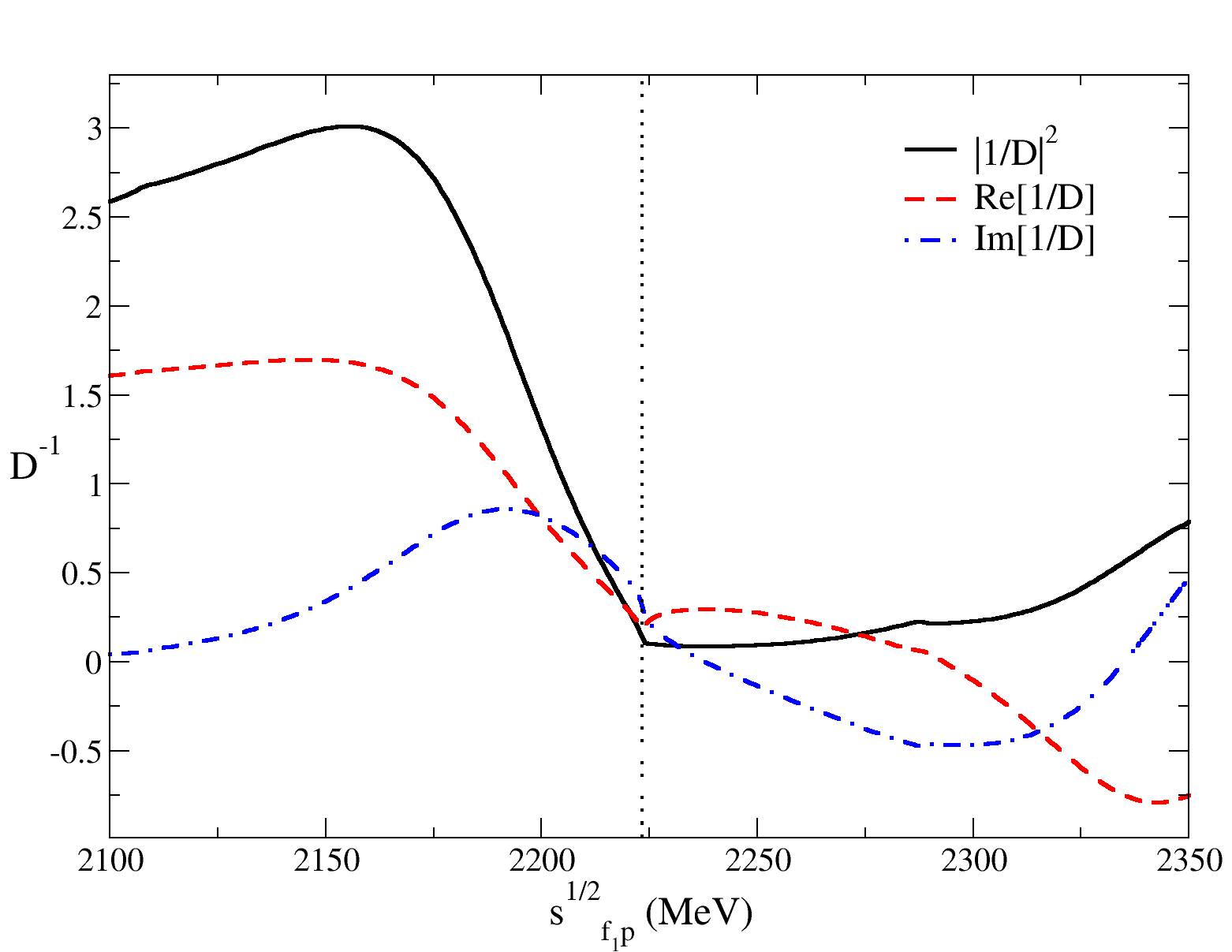}
    \caption{$\text{Re}D^{-1}$, $\text{Im}D^{-1}$ and $|D^{-1}|^2$ as a function of $\sqrt{s}$.}
    \label{fig:denominator}
\end{figure}

It is interesting to see that this peak does not reflect any particular two-body process, but actually comes from the denominator of the amplitude, that is, the factor $D^{-1}=(1-\tilde{t}_1\tilde{t}_2\tilde{G}^2)^{-1}$ in $\tilde{T}$. Indeed, in Fig.~\ref{fig:denominator} we plot the real part, the imaginary part, and the absolute value squared of $D^{-1}$, which show similar characteristics as $\tilde{T}$ in Fig.~\ref{fig:amplitude}.

We should note that if any of the amplitudes $\tilde{t}_i$ were to diverge, $\tilde{T}$ would still have a soft behaviour through cancellations between the numerator and denominator in Eq.~(\ref{eq:Ttotal}). However, it is the factor $D^{-1}$ that reflects the multiple rescattering of the $N$ inside the cluster and provides the resonance structure of $\tilde{T}$.

\begin{figure}
    \centering
    \includegraphics[width=\linewidth]{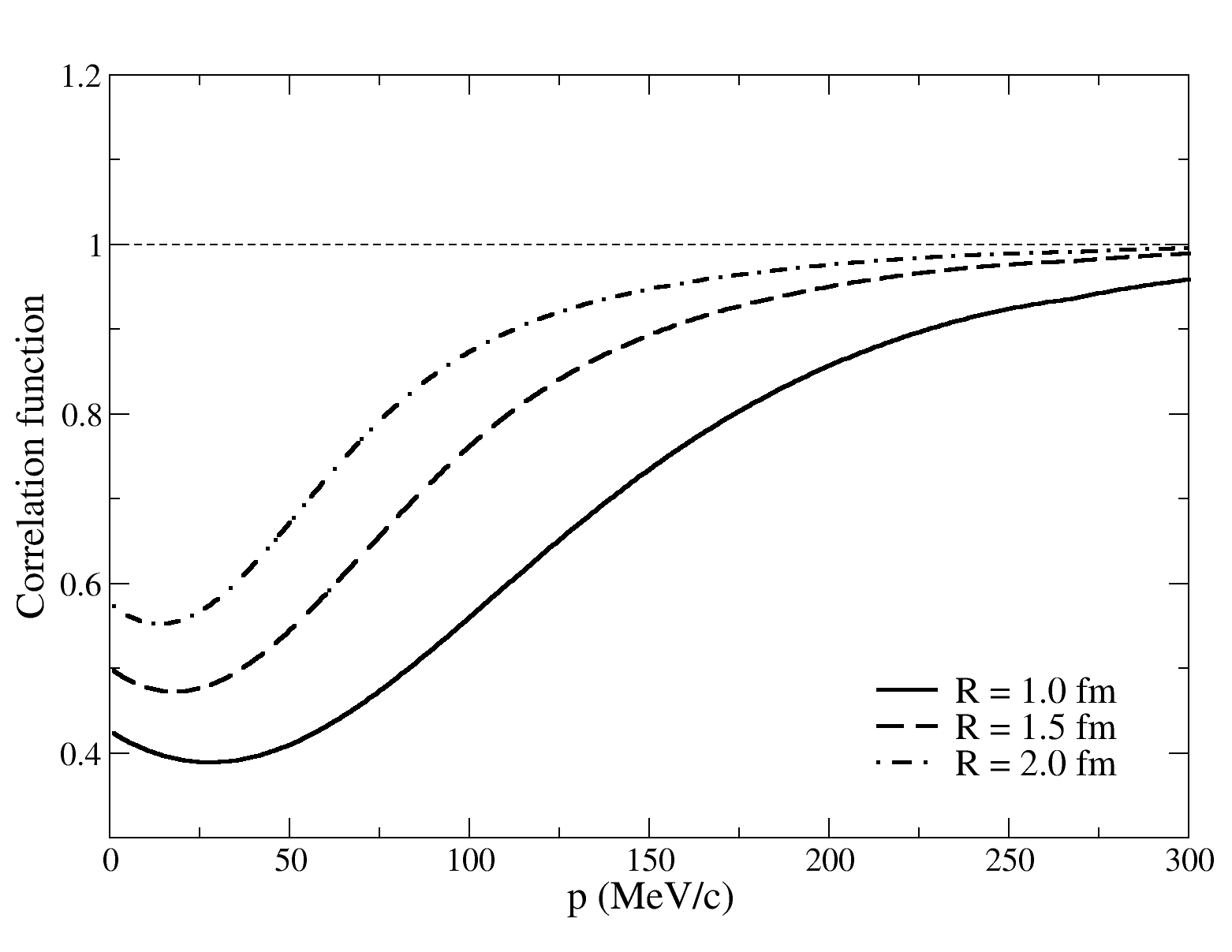}
    \caption{$p~f_1(1285)$ correlation function for different values of $R=1.0,\ 1.5,\ 2.0$ fm.}
    \label{fig:CF}
\end{figure}

Next we plot in Fig.~\ref{fig:CF} the $p~f_1(1285)$ correlation function. We evaluate the CF up to $p=300$ MeV/c ($p$ corresponds to $q_{cm}$ in Eq.~(\ref{eq:qqcm})), refraining to go further up in momentum to prevent entering regions where the FCA does not give reliable results. Indeed, while the bound region and close to threshold the results of the FCA are accurate, it was found in \cite{MartinezTorres:2010ax} that for energies around $200$ MeV above threshold, the FCA already showed clear deficiencies. It is not surprising, since the main assumption in the FCA is that the cluster does not break in the succesive scatterings of the external particle with the components of the cluster. While this is expected to be accurate when the components of the cluster are well bound  ($100$ MeV in the present case) and there is not enough energy of the external particle to break the cluster, this should not be the case when the external particle is energetic enough to break it. In the present case one can go safely around $300$ MeV/c of the relative momentum, which corresponds to $\sqrt{s}$ around $75$ MeV above the $p~f_1(1285)$ threshold.

We see in Fig.~\ref{fig:CF} that the correlation function shows a depletion at low momentum, a common result when one has a resonance of the system below threshold, as seen for instance in \cite{ALICE:2023wjz}, where this shape in the $\Lambda K$ CF is due to the $N^*(1535)$ located right below threshold \cite{Molina:2023jov}, in analogy to the present case. The correlation function increases smoothly and approaches unity at $300$ MeV/c for large source sizes, while for smaller $R$ the convergence occurs further up in momentum.

It would be interesting to compare the predictions made here with the actual experimental results, which are under current analysis at ALICE \cite{Otoninfo}.

\subsection{Uncertainty estimate}
\begin{figure}
    \centering
    \includegraphics[width=\linewidth]{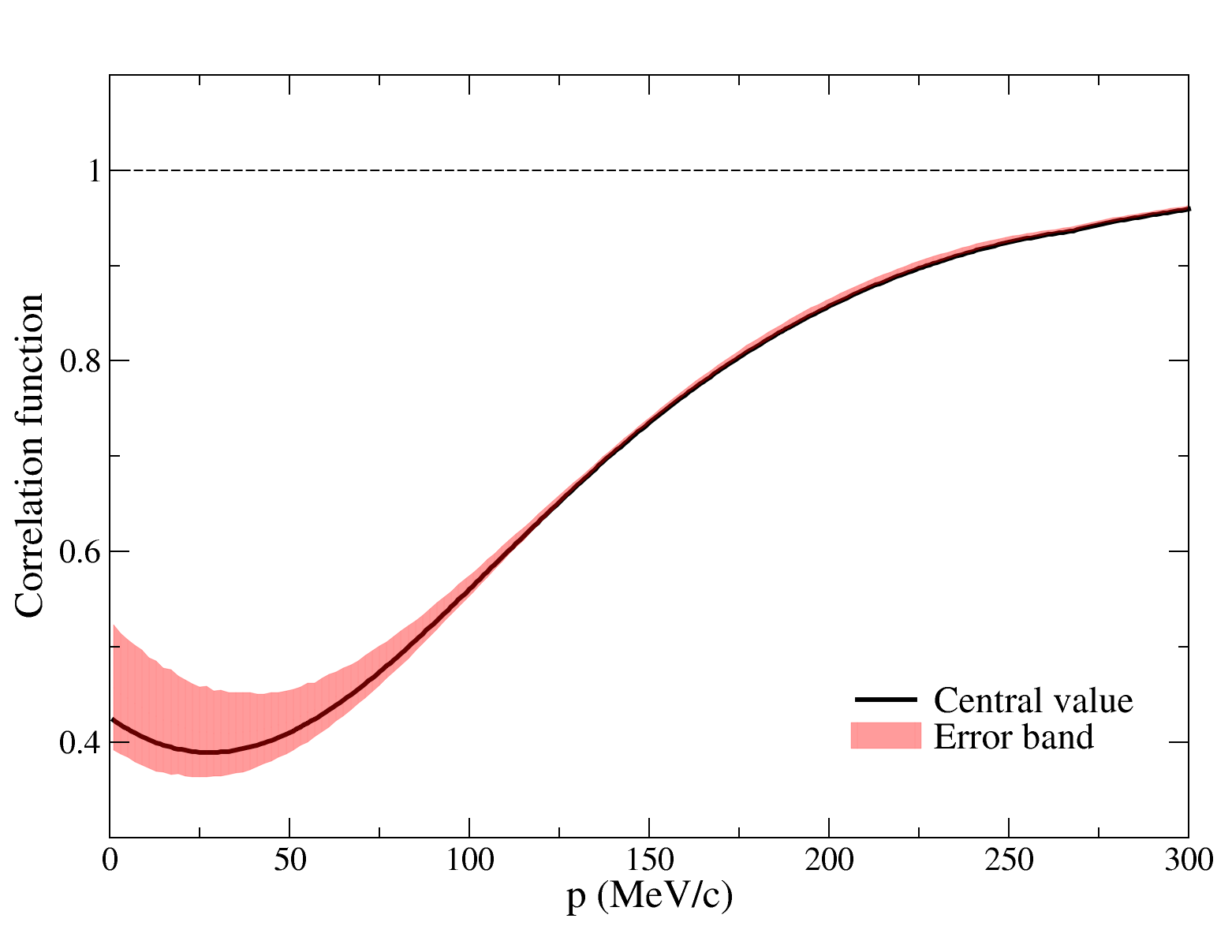}
    \caption{$R=1.0$ fm CF with its corresponding error band estimation (see text for more details). The black line is the one obtained with the parameters used to generate the solid curve in Fig.~\ref{fig:CF}.}
    \label{fig:CF_erband}
\end{figure}
It is interesting to provide a range of uncertainty to our results. We look into amplitudes of Appendix~\ref{ap:apA} for uncertainties in the input and we find that, by large, the biggest uncertainties come from Eq.~(\ref{BW_NKst}), approximating the $T_{N\bar{K}^*,N\bar{K}^*}$ amplitude. We take theoretical uncertainties in the coupling, and the $\Lambda(1800)$ mass and width errors from the PDG compilation. We take
\begin{equation}
 g_{N\bar{K}^*}^{I=0} \in [3.30, 4.00] \, ,   
\end{equation}
where the first number corresponds to Ref.~\cite{Oset:2010tof} and the second to Ref.~\cite{Garzon:2012np}, and 
\begin{eqnarray}
 M_{\Lambda(1800)} &\in& [1750, 1850] \text{ MeV} \, , \nonumber \\
 \Gamma_{\Lambda(1800)} &\in& [150, 250] \text{ MeV} \, .
\end{eqnarray}
We choose gaussian distributed random values of these parameters and evaluate the correlation function for $R=1$~fm of Fig.~\ref{fig:CF} for comparison. The obtained error band, removing the 16\% higher and lower values to have 68\% confidence level, is shown in Fig.~\ref{fig:CF_erband}. This also serves to give uncertainties in the peak position and width of the 3 body state as 
\begin{eqnarray}
     M &=& 2185 \pm 7 \text{ MeV} \, , \nonumber \\
     \Gamma &=& 77 \pm 18 \text{ MeV} \, ,
\end{eqnarray}
and the scattering length and effective range as
\begin{eqnarray}
    a_0 &=& (1.06 \pm 0.12) + i(-0.63 \pm 0.09) \text{ fm} \nonumber \\
    r_0 &=& (1.23 \pm 0.65) + i (1.46 \pm 0.51) \text{ fm} \, .
\end{eqnarray}

\subsection{Comparison with Koonin-Pratt using the effective range expansion}

\begin{figure}
    \centering
    \includegraphics[width=\linewidth]{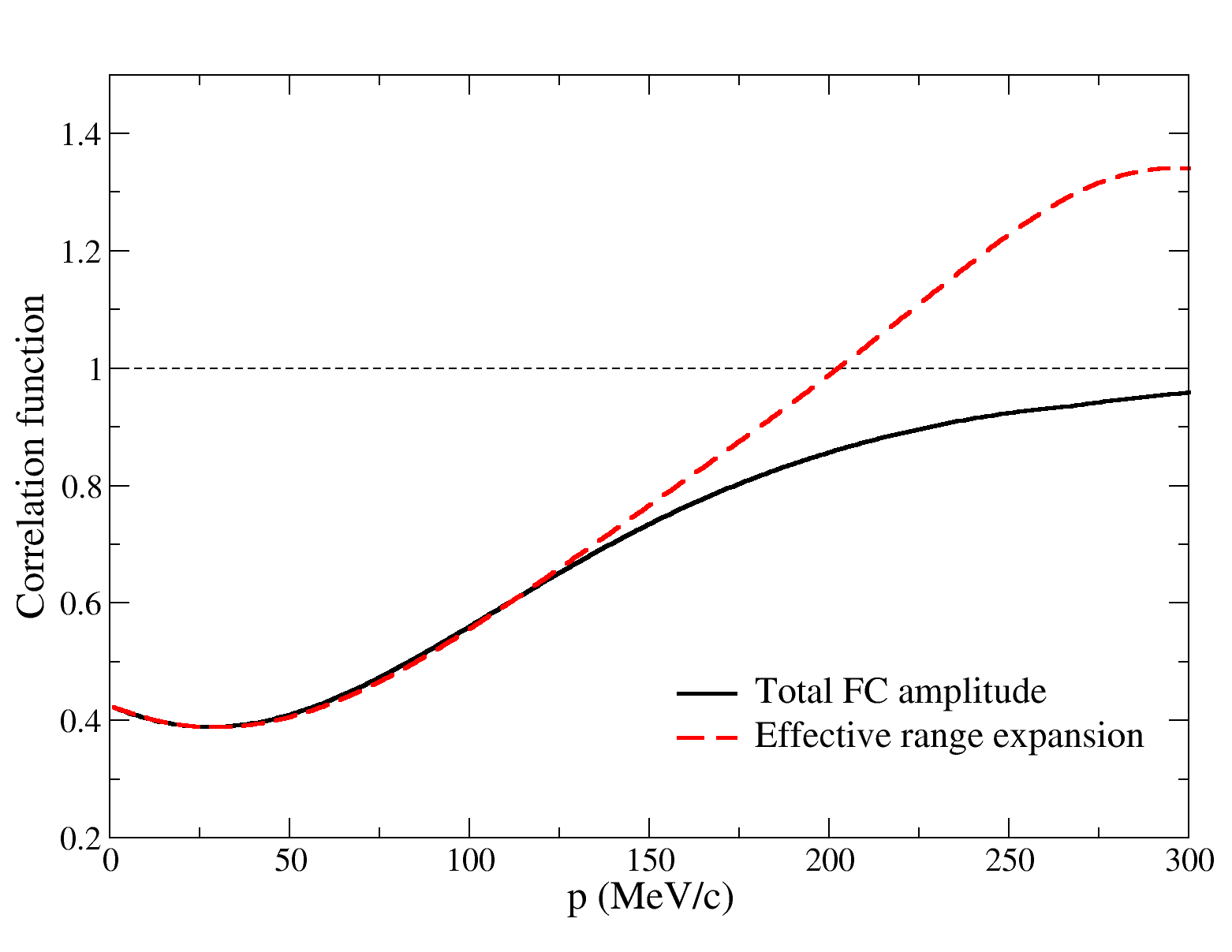}
    \caption{Comparison between the CF using the total $\tilde{T}$ amplitude and its effective range expansion, for $R=1.0$ fm.}
    \label{fig:CF_Exp}
\end{figure}

Using Eqs.~(\ref{eq:a0r0}) we calculate the scattering length and effective range, and we find
\begin{eqnarray}
    a_0 &=& 1.04 - i0.57\text{ fm} \nonumber \\
    r_0 &=& 1.17 + i1.16\text{ fm}
\end{eqnarray}
In Fig.~\ref{fig:CF_Exp} we compare the obtained results with those of the Koonin-Pratt formula \cite{Koonin:1977fh}, using for $\tilde{T}$ of Eq.~(\ref{eq:CF}) the effective range expansion of Eq.~(\ref{eq:effectiverangeexpansion}) with the obtained scattering parameters.

We observe that our full curve compares very well with the effective range expansion up to $p\sim125$ MeV. This means that our approach is much richer, generating terms in the effective range expansion beyond the $q^2$ term. Note that our curve converges faster to unity, and does not have the bump that the effective range approximation produces.

\section{Results assuming a nonmolecular picture of the $f_1(1285)$}
The compositeness of a state measures the probability that this state is a molecular state of the given components, being in the $f_1(1285)$ case the $K\bar{K}^*$ or $\bar{K}K^*$. This probability is commonly expressed as $-g^2 \partial G/\partial s$ \cite{Gamermann:2009uq,Hyodo:2011qc,Hyodo:2013nka,Sekihara:2014kya}, with $g$ the coupling of the state to these components and $G$ the loop function for the propagation of the components, see Appendix~\ref{ap:apA}. By construction of the molecular state for the $f_1(1285)$, this magnitude is unity. Yet, we find it instructive to see if we could learn something about the nature of the $f_1(1285)$ from the knowledge of the correlation function. For this purpose we shall assume that the correlation function has been measured and one has the results of Fig.~\ref{fig:CF} for $R=1$~fm. Then we assume that the $f_1(1285)$ is an ordinary state and try to see if it can reproduce this correlation function and which repercussions it has. Bearing this in mind, we consider that $f_1(1285)$ is an elementary particle which interacts with the proton and we assume a potencial 
\begin{equation}
    V=\frac{C}{4f^2}\frac{m_p+E_p}{m_p}(\sqrt{s}-m_p) \, , \,\,\, f=93\text{ MeV}\, ,
\label{eq:nonmolkernel}    
\end{equation}
where the mass and energy of the proton is denoted by $m_p$ and $E_p$, while $\sqrt{s}$ represents the total energy of the $f_1p$ system in the center-of-mass (CM). The $C$ coefficient stands for the unkown interaction coupling between $f_1$ and $p$, which is fitted to a synthetic data set generated from the $R=1$~fm $f_1p$ CF in Fig.~\ref{fig:CF}. This particular recipe for the interaction kernel allows a comparison with ordinary potentials of the chiral unitary approach \cite{Oset:1997it}. The scattering matrix is given by 
\begin{equation}
    T=[1-VG]^{-1}V \, ,
\label{eq:tmatrix}
\end{equation}
with the loop function $G$ defined as
\begin{eqnarray}
    \label{eq:determine_qcut}
 G(\sqrt{s})&=&\int_{|\vec{q}|<q_{\text{max}}}\frac{d^3q}{(2\pi)^3}\frac{m_p}{2\omega_{f_1}(\vec{q})E_p(\vec{q})}  \nonumber \\
 & & \times \frac{1}{\sqrt{s}-\omega_{f_1}(\vec{q})-E_p(\vec{q})+i\epsilon},
\end{eqnarray}
with $E_p=\sqrt{m_p^2+q^2}$ and $\omega_{f_1}=\sqrt{m_{f_1}^2+q^2}$, being $m_p$ and $m_{f_1}$ the corresponding proton and $f_1(1285)$ masses.
\begin{figure}
    \centering
    \includegraphics[width=\linewidth]{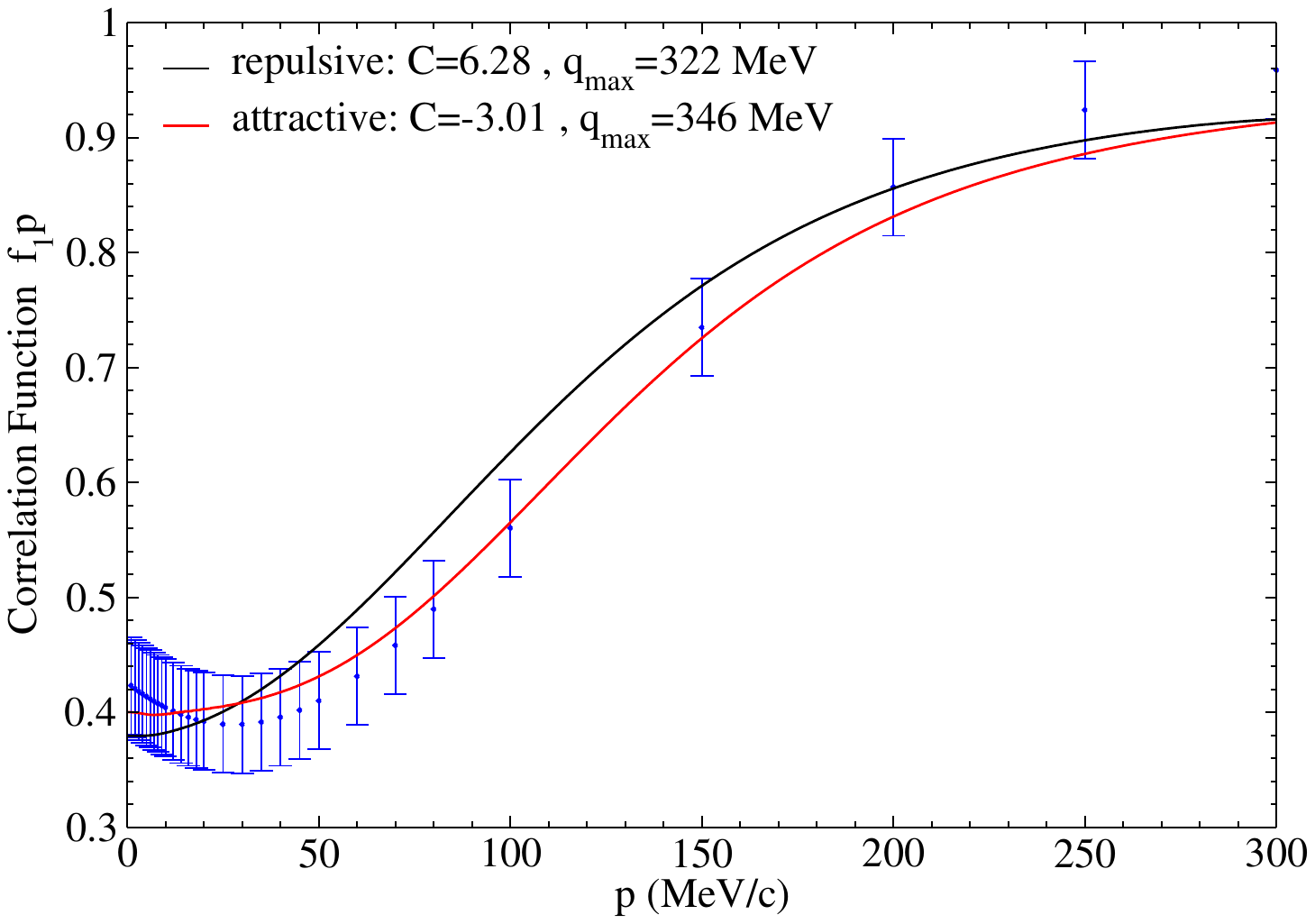}
    \caption{The synthetic data points and the associated error bars are generated from the $R=1$~fm of Fig.~\ref{fig:CF} (see the text for more details). The red and black solid lines represent the $p f_1$ CF for the attractive and repulsive solutions obtained from the fitting procedure within the assumption of a nonmolecular $f_1$ state.}
    \label{fig:nonmolecf1}
\end{figure}

In order to generate the pseudodata, we assume the same error for each point, as shown in Fig.~\ref{fig:nonmolecf1}. We take a set of points inhomogeneously distributed with a more populated low energy range, since this region is the one that interests us the most. The parameters $C$ and $q_{\text{max}}$, entering in Eq.~(\ref{eq:nonmolkernel}) which is incorporated in the computation of the new $T$-matrix of Eq.~(\ref{eq:tmatrix}), are fitted to the synthetic data through Eqs.~(\ref{eq:CF})-(\ref{eq:loopCF}). We find two possible solutions. One corresponds to a repulsive potential, with 
\begin{equation}
C=6.28\, ,  \,\,\,\, q_{\text{max}}=322 \text{MeV} \, .
\end{equation}
Using Eqs.~(\ref{eq:a0r0}) we can also obtain the scattering parameters $a_0$ and $r_0$, which are now real as
\begin{equation}
a_0 = 2.91 \text{ fm} \, ,  \,\,\,\, r_0 = 40.52 \text{ fm} \, .
\end{equation}
Note that while $a_0$ is of the order of the one found before, $r_0$ becomes abnormally large. Certainly, this solution, which follows the trend of ordinary repulsive potentials (see \cite{Liu:2023uly}), does not generate any bound state.
We find also an attractive solution that leads to a bound state. The values of the corresponding parameters are 
\begin{equation}
C=-3.01\, ,  \,\,\,\, q_{\text{max}}=346 \text{MeV} \, ,
\end{equation}
which provide as scattering parameters
\begin{equation}
a_0 = 2.37 \text{ fm} \, ,  \,\,\,\, r_0 = 0.69 \text{ fm} \, .
\end{equation}
The values of $a_0$, $r_0$ are now more in line with those obtained from the molecular picture, and the behavior of $p f_1$ CF at small values of $p$ would make it preferable to the repulsive solution (see red and black curves in Fig.~\ref{fig:nonmolecf1}). By looking at $V^{-1}-G$ below threshold, we also find that this attractive potential generates a bound state $p f_1$ with a binding energy of about $9$~MeV when $V^{-1}-G=0$. This output is in contrast with the state obtained from the molecular picture of around $40$~MeV. We can also say something about the width of this state. In the elementary particle picture we would assume that the width corresponds to the natural width of the $f_1(1285)$ of about $23$~MeV. Instead the width of the three body state assuming the molecular picture of the $f_1(1285)$ is of the order of $70$~MeV. These features could be investigated experimentally. We propose to measure for instance one of the $f_1(1285)$ decay channels for the case that the $f_1$ would be an elementary state, like $\bar{K} \pi K$ channel, which is used to identify the $f_1(1285)$ experimentally \cite{Otoninfo}. This, in addition to a proton, would be a signal of the three body state. Conversely, for the case where the $f_1$ state is of molecular nature, we suggest to investigate the $\Sigma \pi K^*$ decay channel, which would come from $\bar{K}N$ to $\Sigma \pi$ from the $p \bar{K} K^*$ original state. The prediction of the two models are sufficiently different to decide in favor of one or the other picture.   

\section {Conclusions}
We have addressed the novel problem of determining the correlation function for the interaction of a stable particle with a resonance state. We have chosen a particular case, the interaction  of a proton with the $f_1(1285)$ state, taking advantage that the analysis of the data for this problem is currently underway in the ALICE collaboration. The correlation function is obtained with a formalism that requires the scattering matrix of the proton with the $f_1(1285)$ state. In this case, one can think of the resonance $f_1(1285)$ as a dynamically generated state, which comes from the $K^*\bar{K}$ and $\bar{K}^*K$ interaction, leading to a bound state of these components. To evaluate the $p~f_1(1285)$ scattering matrix, one has to apply some nuclear physics technique, of $p$ interaction with a bound state of $K^*\bar{K}$ and $\bar{K}^*K$. We have chosen to face this problem using the fixed center approximation, which is quite reliable, particularly when one has a relatively large binding of the molecular components, as is the case for the $f_1(1285)$. We find that the $p~f_1(1285)$ interaction produces a resonant state below the threshold at $2182$ MeV, with a binding energy of $40$ MeV  and a width of about $78$ MeV. 

With the evaluated $p~f_1(1285)$ scattering matrix we compute the   $p~f_1(1285)$ correlation function, and we find that it has a depletion at small values of the proton momentum, and increases smoothly towards unity. It will be very interesting to compare these theoretical results with the experiment when it is finalized. Because this is a new brand of experiment, the value of the size of the source function is not known, but we recall that this is a magnitude that can be obtained from the experiment in the many checks that have been done with a model independent analysis of data. Because of this, we have evaluated the correlation function for different values of R, the radius of the source function.  

The formalism developed here is applied for the first time to evaluate the correlation function of a stable particle with a resonance state of molecular nature and can be applied to many other cases. Since the results are intimately tied to the molecular nature of the resonance chosen, the results for the correlation function are bound to provide valuable information on the nature of many resonances that can qualify for this type of structure. At the same time the experiment will be providing information on the likely formation of three body bound states, as has occurred in the present case. We also discussed that an analysis of the correlation function and invariant mass distributions of the likely decay channels of the three body system found can shed some light on the nature of the $f_1(1285)$ in the present case. One anticipates much progress in hadron physics from the development of this line of work. 

\begin{acknowledgments}

The authors are very grateful to N. Ikeno for the fruitful discussions and her careful reading of the manuscript. This work is partly supported by  the Spanish Ministerio de Economia y Competitividad (MINECO) and European FEDER funds under Contracts No. FIS2017-84038-C2-1-P B, PID2020- 112777GB-I00, and by Generalitat Valenciana under contract PROMETEO/2020/023. This project has received funding from the European Union Horizon 2020 research and innovation
programme under the program H2020- INFRAIA2018-1, grant agreement No. 824093 of the STRONG-2020 project. This work is supported by the Spanish Ministerio de Ciencia e Innovación (MICINN) under contracts PID2020-112777GB-I00, PID2023-147458NB-C21 and CEX2023-001292-S; by Generalitat Valenciana under contracts PROMETEO/2020/023 and CIPROM/2023/59.

\end{acknowledgments}

\appendix
\section{Amplitudes needed in the approach}
\label{ap:apA}

We need the following amplitudes to compute the total $\tilde{T}$ amplitude in Eq.~(\ref{eq:Ttotal}):
\begin{enumerate}
    \item[1)] $\bar{K}N$, $I=0$
    
    We obtain this one using the chiral unitary approach of \cite{Oset:1997it} with the coupled channels $\bar{K}N,\pi\Sigma,\eta\Lambda,K\Xi$. The $T$ matrix is given by
    \begin{equation}
        T=[1-VG]^{-1}V
    \end{equation}
    with $G$ the loop function regularized with a cutoff $q_{max}=630$ MeV, and with
    \begin{equation}
        V_{ij} = -\frac{1}{4f^2}D_{ij}(k^0+k'^0),
    \end{equation}
    where $k^0,k'^0$ are the energies of the initial and final mesons, and $f$ the effective meson decay constant, taken here as $f=1.15f_{\pi}$, with $f_{\pi}=93$ MeV. The coefficients $D_{ij}$ are given in Table 2 of \cite{Oset:1997it}.
    \item[2)] $\bar{K}N$, $I=1$
    
    We follow the same work \cite{Oset:1997it} with the coupled channels $\bar{K}N,\pi\Sigma,\pi\Lambda,\eta\Sigma,K\Xi$. The potential is now given by
    \begin{equation}
        V_{ij} = -\frac{1}{4f^2}F_{ij}(k^0+k'^0),
    \end{equation}
    where the $F_{ij}$ coefficients are given in Table 3 of \cite{Oset:1997it}.
    \item[3)] $KN$, $I=0$, $I=1$
    
    Here we only have the $KN$ channels. The interaction is given in \cite{Oset:1997it} as
    \begin{equation}
        V^{(I)} = -\frac{1}{4f^2}L^{(I)}(k^0+k'^0),
    \end{equation}
    where $I$ stands for the isospin and $L^{(0)}=0$ and $L^{(1)}=-2$ are obtained from Table 4 of \cite{Oset:1997it}. This interaction is zero for $I=0$ and repulsive for $I=1$.
    \item[4)] $\bar{K}^*N$, $I=0$
    
    We follow here the work of \cite{Oset:2010tof}, and the coupled channels are now $\bar{K}^*N,\omega\Lambda,\rho\Sigma,\phi\Lambda,K^*\Xi$. The potential is given by
    \begin{equation}
        V_{ij} = -\frac{1}{4f^2}C_{ij}^{I=0}(k^0+k'^0)\vec{\epsilon}\cdot\vec{\epsilon}',
    \end{equation}
    and the coefficients $C_{ij}^{I=0}$ are given in Table 8 of \cite{Oset:2010tof}.

    Here we must make an observation. The approach generates a state $\Lambda^*$ at $1800$ MeV, corresponding to the $\Lambda(1800)$ coupling mostly to $\bar{K}^*N$. However, with $\bar{K}^*N$ having the smallest threshold of those channels, the width is zero (not exactly if the $K^*$ width is considered). A width is obtained if the pseudoscalar-baryon channels are considered as possible decay channels \cite{Garzon:2012np}. We adopt a pragmatic approach here and take an empirical amplitude
    \begin{equation}
        T_{N\bar{K}^*,N\bar{K}^*}^{(I=0)}(\sqrt{s}) = \frac{g_{N\bar{K}^*}^2}{\sqrt{s} - M_{\Lambda^*} + i \Gamma_{\Lambda^*}/2}
     \label{BW_NKst}   
    \end{equation}
    as done in \cite{Bayar:2024sou}, and we take the mass $M_{\Lambda^*} = 1800$ MeV and width $\Gamma_{\Lambda^*}=200$ MeV from the PDG \cite{ParticleDataGroup:2024cfk}, and the coupling from \cite{Garzon:2012np} as $|g_{N\bar{K}^*}^{I=0}|=2.83\sqrt{2}$.
    \item[5)] $\bar{K}^*N$, $I=1$
    
    We have the coupled channels $\bar{K}^*N,\rho\Lambda,\rho\Sigma,K^*\Xi,\phi\Sigma$. The potential is given by
    \begin{equation}
        V_{ij} = -\frac{1}{4f^2}C_{ij}^{I=1}(k^0+k'^0)\vec{\epsilon}\cdot\vec{\epsilon}',
    \end{equation}
    and the coefficients $C_{ij}^{I=1}$ are given in Table 9 of \cite{Oset:2010tof}.
    \item[6)] $K^*N$, $I=0$, $I=1$
    
    Similarly to what is done with $KN$, the potential is given by
    \begin{equation}
        V^{(I)} = -\frac{1}{4f^2}L^{(I)}(k^0+k'^0)\vec{\epsilon}\cdot\vec{\epsilon}',
    \end{equation}
    with $L^{(0)}=0$ and $L^{(1)}=-2$.
\end{enumerate}

In the cases which include the $K^*$ or $\rho$ vectors, we convolve the corresponding $G$ functions with the mass distributions of $K^*$ and $\rho$ to account for the width of these states, as done in \cite{Oset:2010tof}.

\bibliography{apssamp}

\end{document}